COMMENT on **"Altermagnetic and Dipolar Spitting of Magnons in $FeF_2$"**

S. W. Lovesey[1, 2, 3]

[1] ISIS Facility, STFC, Didcot, Oxfordshire OX11 0QX, UK

[2] Diamond Light Source, Harwell Science and Innovation Campus, Didcot, Oxfordshire OX11 0DE, UK

[3] Department of Physics, Oxford University, Oxford OX1 3PU, UK

Conclusions by Sears *et al.* about the dispersion of magnons in the rutile antiferromagnet $FeF_2$ are likely unsound, because compulsory lattice degrees of freedom are omitted [1]. Van Vleck's approach to relaxation times in magnetic solids is valid in light of angular anisotropy in the Fe atomic configuration $3d^6$ [2,3]. Sears *et al.* do not mention lattice degrees of freedom in the construction of a model they confront with neutron time-of-flight spectra using neutrons from a spallation source. Turning back the clock more than half a century, Rainford *et al.* published high-resolution magnon dispersions for $FeF_2$, gathered with slow neutrons from a steady-state reactor [4]. The dispersions are consistent with hybridized magnon-phonon excitations in the region of wavevector-energy space of principal interest for the posited conclusions [1,5] (some experimental data donated by Rainford *et al.* appear in Ref. [5]). Rainford *et al.* achieved exemplary wavevector-energy resolution with a triple-axis spectrometer, whereby focussing aligned the principal axis of the instrument's resolution ellipsoid and a tangent to hybridized excitations [6]. Additional clarification about hybridized excitations in $FeF_2$ awaits neutron spectroscopy and a well-characterized sample exposed to a steady magnetic field, together with neutron polarization analysis [1,7].